\documentstyle[12pt]{article}
\begin{document}
\pagestyle{plain}
\newcommand{\be}{\begin{equation}}
\newcommand{\ee}{\end{equation}}
\newcommand{\bea}{\begin{eqnarray}}
\newcommand{\eea}{\end{eqnarray}}
\newcommand{\vp}{\varphi}
\newcommand{\pr}{\prime}
\newcommand{\sech} {{\rm sech}}
\newcommand{\cosech} {{\rm cosech}}
\newcommand{\psib} {\bar{\psi}}
\newcommand{\cosec} {{\rm cosec}}
\def\vs {\vskip .3 true cm}
\hfil {IP/BBSR/95-111 }\break

\hfil {December'95}\break
\vskip .3 true cm
\centerline {\bf A Class of Exact Solutions for N-anyons}
\centerline {\bf in a N-body potential}
\vs
\centerline {\bf Avinash Khare$^*$}
\centerline { Institute of Physics, Sachivalaya Marg,}
\centerline {Bhubaneswar-751005, India}
\vs
\noindent {\bf Abstract}

A class of exact solutions are obtained for the problem of N-anyons
interacting via the N-body potential

$V (\vec x_1,\vec x_2,...,\vec x_N)$ = $-{e^2\over\sqrt{{1\over
N}\sum_{i<j} (\vec x_i-\vec x_j)^2}}$

Unlike the oscillator case the resulting spectrum is not linear in
the anyon parameter $\alpha (0\leq \alpha\leq 1)$. However, a la
oscillator case, cross-over between the ground states is shown to
occur for N-anyons $(N\geq 3)$ experiencing the above potential.
\vfill
* e-mail : khare@iopb.ernet.in
\eject
By now it is well established [1,2] that in two spatial dimensions
one can have anyonic statistics which interpolates between the
Bose-Einstein and the Fermi-Dirac statistics. Such objects also arise
in 2+1 dimensional field theory as classical solutions of the abelian
Higgs model with a Chern-Simons term [3]. It has been suggested that
anyons may provide mechanism for the fractional quantum Hall effect [4].

In the anyonic quantum mechanical systems, only the problem of two
anyons in various potentials has been solved exactly and as a result
only the second virial coefficient of an anyon gas has been computed
exactly [2]. The exact solution of the N-body problem $(N \geq 3)$
seems to be out of reach. This is rather unfortunate as the nontrivial
braiding effect of anyons is expected to show up only for $N \geq
3$, since only then the 3-body anyonic interaction manifests itself.
As far as I am aware off, to date, only a class of exact solutions
have been obtained in case N-anyons $(N\geq 3)$ experience harmonic
oscillator potential [5] or are in a uniform magnetic field [6]
(which actually is equivalent to the oscillator problem except for a
piece coming from the angular momentum eigenvalue). In both
of these problems, all the known exact solutions are such that the
energy eigenvalue spectrum is linear in the anyon parameter $\alpha$
($0\leq \alpha \leq 1$ and throughout this note $\alpha$ = 0(1) will correspond
to boson (fermion)).

It is clearly of interest to enquire if one can also obtain exact
solutions in case N-anyons are experiencing some other potential and
if in these cases also the energy varies linearly with $\alpha$ or
not. The purpose of this letter is to present one such example. In
particular, I obtain a class of exact solutions in case N-anyons are
interacting via the N-body potential
\be\label{1}
V(\vec x_1,\vec x_2,...,\vec x_N) = - {e^2\over\sqrt{{1\over
N}\sum_{i<j} (\vec x_i-\vec x_j)^2}}
\ee
The interesting point is that unlike the oscillator case, the energy
spectrum here is not linear in $\alpha$. However, a la
oscillator case, these exact solutions include the ground
state of N-bosons but not the ground state of N-fermions
$(N\geq 3)$. We therefore perturbatively calculate the ground state
energy of three anyons near the fermionic statistics and show that
for this potential also there is a cross-over between the ground
states. I show that a similar cross-over must also occur in the case
of N-anyons $(N\geq 4)$.

For simplicity let us first discuss the case of 3-anyons experiencing
the above 3-body potential. After the separation of the center of
mass (which is independent of anyons), the relative problem is best
discussed in terms of the hyper-spherical coordinates
$\rho,\theta,\phi,\psi$ first proposed by Kilpatrick and Larsen [7].
In particular, one can show that the relative Hamiltonian can be
written as [8]
\be\label{2}
H = H^{rad}_0 +{1\over 2\mu\rho^2}(-\Lambda^2+\alpha H_1+\alpha^2 H_2)
\ee
where $\Lambda, H_1$ and $H_2$ only depend on the angular coordinates
$\theta, \phi,\psi$ while $H_0^{rad}$ only depends on the radial
variable $\rho.$ In particular, $-\Lambda^2$ is the Laplacian on the
three dimensional sphere while the anyonic pieces $H_1$ and $H_2$ are
as given in [9]. It is worth emphasizing that such a separation is
always possible so long as the anyons experience a potential which is
a function of $\rho$ alone. Further, such a decomposition also exists
for an arbitrary number of anyons [10]. In particular, for N anyons, the
relative problem is best discussed in terms of $\rho$ and $2N-3$ angles.

In the case of 3-anyons experiencing the 3-body potential (\ref{1}),
the relative radial Hamiltonian $H^{rad}_0$ becomes
\be\label{3}
H_0^{rad} = {-1\over 2\mu} ({\partial^2\over\partial \rho^2}+{3\over
\rho} {\partial\over \partial \rho}) - {e^2\over\rho}
\ee
Let us first obtain the exact eigenvalues and eigenfunctions of three
bosons and fermions in the potential (\ref{1}). They are obtained by
noticing that in that case $\alpha$ can be taken to be zero and further,
the eigenvalue of $-\Lambda^2$ is k(k+2) with k = 0,1,2,.... The resulting
radial equation as obtained from eqs. (\ref{2}) and (\ref{3}) is nothing but
the Schr\"odinger equation for the Coulomb potential in 4-dimensions. In this
way we find that the energy eigenvalues of 3-bosons or 3-fermions in the
3-body potential (\ref{1}) are
\be\label{4}
E_{n',k} = -{\mu e^4\over 2[n'+k+{3\over 2}]^2}
\ee
while the corresponding fermionic (bosonic) eigenfunctions,
$\psi^{(\mp)}_{n',k,\nu,\lambda}$ are given by
\be\label{5}
\psi^{(\mp)}_{n',k,\nu,\lambda} = F^k_{n'}(\rho) Y^{(\mp)}_{k,\nu,\lambda}
(\theta,\phi,\psi)
\ee
Here the normalized angular eigenfunctions
$Y^{(\mp)}_{k,\nu,\lambda} (\theta,\phi,\psi)$ are identical to
those in the harmonic case and have been explicitly written down in
[9] while the normalized (with measure $\rho^3 d\rho)$ radial
eigenfunctions $F^k_{n'}(\rho)$ are given by
\be\label{6}
F^k_{n'}(\rho) = N_{n',2k+2} exp(-y/2) y^k L_{n'}^{2k+2} (y)
\ee
where $L_N^{\alpha}$ is a Laguerre polynomial, $y = 2\sqrt{2\mu\mid
E\mid}\rho$ and $N_{n',2k+2}$ is the normalization constant. It is
worth pointing out that for the 3-boson ground state, $n'=k=0$ and
hence $\in^B_g\equiv  2 E^B_g/\mu e^4 = - 4/9$ while for 3-fermion
ground state $n'=0, k =2$ and hence $\in^F_g = -4/49$. It is also
worth pointing out that the wave function (\ref{5}) is also an
eigenfunction of the angular momentum operator with eigenvalue $\lambda$.

Proceeding in the same way, the eigenvalues of N bosons or N fermions
in the N-body potential (\ref{1}) can be immediately written down.
This is because in that case $-\Lambda^2$ is the Laplacian on the
(2N-3)-dimensional sphere whose eigenvalues are k(k+2N-4) with k =0,
1,2,..., and whose eigenfunctions are generalized spherical
harmonics. The resulting radial Schr\"odinger equation then takes the form
\be\label {8}
\bigg [{\partial^2\over\partial\rho^2} +{2N-3\over
\rho}{\partial\over\partial\rho} + {2\mu e^2\over \rho}
-{k(k+2N-4)\over\rho^2}\bigg ] F(\rho) = - 2\mu E F(\rho)
\ee
This equation is easily solved and the resulting energy eigenvalues
for N bosons or N fermions experiencing the N-body potential (\ref{1})
are given by $(k,n' =0,1,2,...)$
\be\label{9}
\in_{n',k} = -{1\over[n'+k+N-{3\over 2}]^2}
\ee
while the corresponding unnormalized radial eigenfunctions $F(\rho)$
are given by
\be\label{10}
F^k_{n'} (\rho) = exp(-y/2) y^k L_{n'}^{2N+2k-4} (y)
\ee
where as before $y = 2\sqrt{2\mu\mid E\mid}\rho$.

Let us now turn to the exact solutions of the N anyon problem in the
presence of the N-body potential (\ref{1}). On using the fact that
(i) only the angular part of the Hamiltonian is affected due to the anyons
(ii) the angular part is independent of the radial potential
$V(\rho)$ between the anyons (iii) the radial equation for N bosons, N
fermions and N anyons is same but for the coefficient of the
${1\over\rho^2}$ term, one can immediately write down a class of
exact solutions for N anyons experiencing the N-body potential
(\ref{1}). In particular, on using the exact solutions for  N
anyons in the oscillator potential [5] we find that the exact energy
eigenvalues in our case are
\be\label{11}
\in_{n',\lambda}(\alpha) = -{1\over [n'+\mid\lambda - {N(N-1)\over
2}\alpha\mid + N -{3\over 2}]^2}
\ee
where $\lambda$ is the eigenvalue of the angular momentum operator.
The corresponding eigenfunctions are
\be\label{12}
\psi_{n',\lambda} = exp [ i\lambda \sum_{i<j}\theta_{ij}
]e^{-y/2}\Pi_{i<j} \mid\vec x_{ij}\mid ^{\mid\lambda-\alpha\mid}
L^{a}_{n'} (y)
\ee
where $y = 2\sqrt{2\mu\mid E\mid}\rho, \ a = 2N-4+2\mid\lambda
-{N(N-1)\over 2}\alpha\mid, \ \vec x_{ij} =\vec x_i-\vec x_j$ and
$tan\theta_{ij} ={(y_i-y_j)\over (x_i-x_j)}$.

It is worth pointing out that for N = 2, the expression as given by
eq.(\ref{11}) gives the complete spectrum [11]. For $N\geq 3$
however, it does not give the complete spectrum. For example, for
N = 3, the three fermion ground state is missing from these exact
solutions (the three fermion ground state energy $\in^F_g = -{4\over
49}$ which is not included in the expression (\ref{11})). The three
boson ground state which corresponds to $n' =\lambda =\alpha =0 $ (
and N = 3) has energy $\in^B_g = - 4/9$ and it interpolates to the
fermionic state with $\in^F = - 4/81$ which is an excited
state. Thus, as in the oscillator case [9], a level-crossing has to
occur for the true ground state of the 3-anyon system. Infact, such a
crossing must also occur for any N $( \geq 3)$. This is because the
exact N boson ground state interpolates to the fermionic state with
an eigenstate of angular momentum L with eigenvalue $-N(N-1)/2$ (see
eq.(\ref{11})). On the other hand, the fermionic ground state is
obtained by filling the one particle levels from bottom to top. One
can show that the fermionic ground state always has  a total angular
momentum $\mid L\mid$ less than ${N(N-1)\over 2}$ (for $N > 2$) [12].
We thus conclude that a la oscillator case, even in the case of the
N-body potential (\ref{1}), there must be a ground state cross-over
at some value of $\alpha$.

What is the nature of the missing states in the N-anyon spectra ?
We now show
that in our case, whereas for the exact solutions
$(-\in)^{-1/2}$ is linear in $\alpha$, for all the missing solutions
$(-\in)^{-1/2}$ will have nonlinear dependence on $\alpha$. Let us first recall
that in the case of the oscillator potential, all those states for which
energy varies linearly with $\alpha$ are known analytically [5]. Further, it is
also known that there are several missing states  whose energy varies
nonlinearly with $\alpha$. For N =3,4 the energy of the low lying "nonlinear
states" has been estimated by using numerical and perturbative techniques
[13,14]. We now show that we can borrow these oscillator results and obtain
the energies of the missing states in the case of the potential (\ref{1}).
The point is that as argued above, only the angular part of the Hamiltonian
is affected due to anyons [10] and this part is identical for both the
oscillator
and our potential (\ref{1}) . Secondly, the only effect of the angular part is
to affect the coefficient of the $1/\rho^2$ term in the radial Schr\"odinger
equation. For example, in eq. (\ref{8}), instead of $k(k+2N-4)/\rho^2$ we
would have $\beta (\beta +2N-4)/\rho^2$ where $\beta$ need not
necessarily be an integer and would in general be a complicated function of
$\alpha$. As a result, the energy eigenvalues of the missing states would again
be given by eq. (\ref{9}) but with k replaced by $\beta$. The corresponding
oscillator radial Schr\"odinger equation is also given by eq. (\ref{8}) but
with k replaced by $\beta$ and $2\mu e^2/\rho$ replaced by
$-\mu^2 \omega^2 \rho^2$. As a result, the corresponding oscillator energy
eigenvalues are given by
\be \label{13}
\in^{osc}_{n'} (\alpha) \equiv E_{n'}^{osc} (\alpha)/\omega = (2n'+\beta
+N-1/2)
\ee
One can therefore immediately eliminate $\beta$ from the two eqs. (\ref{13})
and (\ref{9}) (with k replaced by $\beta$) and obtain a general relation
between the eigenvalues of our potential (\ref{1}) and the oscillator
potential given by
\be \label{14}
\in_{n'} (\alpha) = - {1 \over [\in_{n'}^{osc} (\alpha) -n'-1/2 ]^2}
\ee

We can therefore immediately borrow all the known results about the missing
nonlinear states in the oscillator case and obtain corresponding conclusions
in our case. For example, whereas for all the analytically known states
$(-\in)^{-1/2}$ changes by $\pm N(N-1)/2$
for the missing states the energy will change by ${N(N-1)\over 2} -2,
{N(N-1)\over 2} -4,... -[{N(N-1)\over 2} -2]$ as one will go from bosons to
fermions. For example, in the 3-anyon case, $(-\in)^{-1/2}$ changes by $\pm 3$
in the case of the exactly known solutions while it will change by $\pm 1$ in
the
case of the missing states as one will go from bosons to fermions. In
particular,
the 3-fermion ground state at $\in = -4/49$ will interpolate to the bosonic
state
at $\in = -4/81$ and near the fermionic end, the energy of the corresponding
anyonic state is given by [10]
\be \label{15}
\in = - {1\over [3.5 +1.29(1-\alpha)^2]^2}
\ee
On the other hand, the anyonic state starting from the bosonic ground
state at $\in = - 4/9$ is given by (see eq.(\ref{11}))
\be\label{16}
\in = - {4/9\over (1+2\alpha)^2}
\ee
The two curves cross at $\alpha = 0.71$. It is a curious numerical
fact that for both the oscillator [9,13]  and our N-body case, the cross-over
occurs at almost the same point.

Finally it is worth pointing out that the degeneracy of the exact energy levels
coming from the angular part is same for both the oscillator and our
N-body potential (note the same factor $\mid\lambda -{N(N-1)\over
2}\alpha \mid$ occurs in both the cases). This will infact be true for any
anyon potential which only depends on $\rho$. However, the
degeneracy coming from the radial part is different in the two cases
since whereas in the oscillator case one has the factor
$2n'+\mid\lambda- {N(N-1)\over 2}\alpha\mid$, in our case the corresponding
factor is
$n'+\mid\lambda - {N(N-1)\over 2}\alpha\mid$. As a result, compared to
the oscillator case, here the degeneracy is much more. In particular,
for a given
energy, both even and odd angular momentum states are
present in general in the spectra. As a result, if one plots $(-\in)^{-1/2}$
as a function of $\alpha$, then one will find that one will not
only have those levels which are present in the oscillator spectrum
but there will be few extra states in our case which are not there in the
oscillator case. For example, in the 3-anyon case we have an extra
state for which $(-\in)^{-1/2}$ changes linearly from 1.5 to 4.5 as one goes
from
the bosonic to the fermionic end (see for example Fig. 1 of [10] for the low
lying 3-anyon spectrum in the oscillator potential).\\

Are there other potentials for which a class of exact N-anyon eigen
states can be found? We believe that the answer is no since only the
Coulomb and the oscillator problems are analytically solvable in N
dimensions $(N\geq 2)$. All other potentials are atbest
quasi-exactly solvable and hence for a given potential, eigenstates
could be analytically obtained for atbest some specific values of
the angular momentum $\lambda$.

This work raises several issues like number of crossings in the ground state
for N anyons [15], possible supersymmery for N anyons [16],
pseudo-integrability of the N-anyon problem [17], solutions in the presence
of the uniform magnetic field [18], scattering
solutions etc. which need to be discussed carefully in the context of the
potential (\ref{1}). Some of these issues as well as details of
this work will be published elsewhere [19].
\vfill
\eject
\noindent {\bf References}

[1] J.M. Leinaas and J. Myrheim, Nuovo Cimento {\bf 37B} (1977) 1;
G. A. Goldin, R. Menikoff and D.F. Sharp, J. Math. Phys. {\bf 21}
(1980) 650; F. Wilczek, Phys. Rev. Lett. {\bf 49} (1982) 1664.

[2] For a recent review see A. Khare, Current Science {\bf 61} (1991) 826;
A. Lerda, {\it Anyons}, Lecture Notes in Physics {\bf m14},
Springer-Verlag (1992).

[3] S.K. Paul and A. Khare, Phys. Lett. {\bf B174} (1986) 420; {\bf
B182} (1987) E415.

[4] R.B. Laughlin, Phys. Rev. Lett. {\bf 50} (1983) 1395.

[5] Y.S. Wu, Phys. Rev. Lett. {\bf 53} (1984) 111; {\bf 53} (1984)
1028E; C. Chou, Phys. Lett. {\bf A155} (1991) 215; A.
Polychronakos, Phys. Lett. {\bf B264} (1991) 362; R. Basu, G. Date
and M.V.N. Murthy, Phys. Rev. {\bf B46} (1992) 3139.

[6] G.V. Dunne, A. Lerda, S. Sciuto and C.A. Trugenberger, Nucl. Phys.
{\bf B370} (1992) 601; K. Cho and C. Rim, Ann. Phys. {\bf 213} (1992)
295.

[7] J.E. Kilpatrick and S.Y. Larsen, Few-Body Syst. {\bf 3} (1987) 75.

[8] J. McCabe and S. Ouvry, Phys. Lett. {\bf b260} (1991) 113.

[9] A. Khare and J. McCabe, Phys. Lett. {\bf B269 } (1991) 330.

[10] M. Sporre, J.J.M. Verbaarschot and I. Zahed, Nucl. Phys. {\bf
B389} (1993) 645; J. Grundberg, T.H. Hansson, A. Karlhede and E.
Westerberg, Phys. Rev. {\bf B44} (1991) 8373.

[11] J. Law, M.K. Srivastava, R.K. Bhaduri and A. Khare, J. Phys. {\bf A25}
(1992) L183; A. Comtet and A. Khare, Orsay Report IPNO/TH-91-76 (1991)
unpublished; R. Chitra, C.N. Kumar and D. Sen, Mod. Phys. Lett. {\bf A7}
(1992) 855.

[12] A. Khare, J. McCabe and S. Ouvry, Phys. Rev. {\bf D46} (1992) 2714.

[13] M.V.N. Murthy, J. Law, M. Brack and R.K. Bhaduri, Phys. Rev.
Lett. {\bf 67} (1991) 1817; M. Sporre, J.J.M. Verbaarschot and
I. Zahed, Phys. Rev. Lett. {\bf 67} (1991) 1813; C. Chou, Phys. Rev.
{\bf D44} (1991) 2533, {\bf D45} (1992) 1433E.

[14] M. Spoore, J.J.M. Verbaarschot and I. Zahed, Phys. Rev. {\bf B46}
(1992) 5738.

[15] R. Chitra and D. Sen, Phys. Rev. {\bf B46} (1992) 10 923.

[16] D. Sen, Phys. Rev. Lett. {\bf 68} (1992) 2977; Phys. Rev. {\bf D46}
(1992) 1846.

[17] G. Date and M.V.N. Murthy, Phys. Rev. {\bf A48} (1993) 105.

[18] A. Vercin, Phys. Lett. {\bf B260} (1991) 120; J. Myrheim, E. Halvorsen
and A. Vercin, Phys. Lett. {\bf B278} (1992) 171.

[19] A. Khare, Under Preparation.

\end{document}